\documentstyle[12pt,epsfig]{article}

\newlength{\dinwidth}
\newlength{\dinmargin}
\def\lapproxeq{\lower .7ex\hbox{$\;\stackrel{\textstyle<}{\sim}\;$}}
\def\gapproxeq{\lower .7ex\hbox{$\;\stackrel{\textstyle>}{\sim}\;$}}

\def\q/{{\bf q}}
\def\k/{{\bf k}}
\def\ka/{{\bf k_1}}
\def\kb/{{\bf k_2}}
\def\kp/{{\bf k'}}
\def\b/{{\bf b}}
\def\mb/{{\bf \vert b \vert}}
\def\ba/{{\bf b_1}}
\def\bap/{{\bf b'_1}}
\def\bb/{{\bf b_2}}
\def\bbp/{{\bf b'_2}}
\def\r/{{\bf r}}
\def\R/{{\bf R}}
\def\F/{{\tilde{F}}}
\def\T/{\hat{\rm T}}
\def\alp/{\overline{\alpha}_S}

\begin{document}
\titlepage
\begin{flushright}
MC-TH-97/01 \\
LU TP 97-01 \\
hep-ph/9703225 \\
March 1997 \\
\end{flushright}

\begin{center}
\vspace*{2cm}
{\Large \bf 
Diffusion and the BFKL Pomeron
} \\
\vspace*{1cm}

J.R.~Forshaw$^1$ and P.~J.~Sutton$^2$

\vspace*{0.5cm}

$^1$ Department of Physics and Astronomy, 
University of Manchester,\\
Manchester, M13 9PL, England.

$^2$ Department of Theoretical Physics,
Lund University, \\
S\"olvegatan 14 A, S-223 62 Lund, Sweden.
\end{center}

\vspace*{5cm}
\begin{abstract}
We study the high energy behaviour of elastic scattering amplitudes
within the leading logarithm approximation. In particular, we
cast the amplitude in a form which allows us to study the internal 
dynamics of the BFKL Pomeron for general momentum transfer. 
We demonstrate that the momentum transfer
acts as an effective infrared cut-off which ensures that the dominant 
contribution arises from short distance physics.
\end{abstract}

\newpage


\noindent {\large \bf 1.  Introduction}

We are interested in the study of elastic scattering in the Regge limit
where $s \gg -t$. We might expect to be able to use the methods of
perturbative QCD if the typical distances involved in the interaction
are small.  Whilst this may be true for some elastic scattering processes 
(for example,
two highly virtual photons at large $t$) it is certainly not the case 
for others (for example, a pair of protons at $t \simeq 0$). 
It is therefore necessary to check the
self consistency of a perturbative calculation by investigating the
typical distances involved in the interaction. If these turn out to be
large then, even if the calculation is infrared finite, there is little
justification for using a perturbative approach. In this paper, we
consider the elastic scattering of generic colourless states via
BFKL Pomeron \cite{bfkl} exchange. We formulate the amplitude 
in such a way that we can
readily identify the typical distances involved in the exchange. 

For elastic scattering at $t=0$, the typical transverse
momenta of the gluons which constitute the Pomeron are determined
primarily by the sizes of the external particles. By studying the
scattering of small size objects one therefore expects that perturbative
QCD is valid. However, one must also take into account the diffusion
properties characteristic of the BFKL exchange, namely as one moves away in
rapidity from the external particles the width of the transverse
momentum distribution broadens. 
This means that at sufficiently high
centre of mass energies one is destined to pick up a large contribution
from long distance effects. It is the main purpose of this paper
to demonstrate that for $t \neq 0$, the situation changes rather
dramatically. The scale $\vert t \vert $ 
effectively acts as an infrared cut-off, ensuring
that the essential physics comes from the region where all the momenta
are bigger than $\sim \sqrt{-t}$. As a by--product we derive general
formulae, which are useful in computing elastic scattering
amplitudes in the Regge limit.


\vspace*{4mm}
\noindent {\large \bf 2. The BFKL Pomeron}

\begin{figure}
\begin{center}
\leavevmode
\hbox{\epsfxsize=1.8 in 
\epsfbox{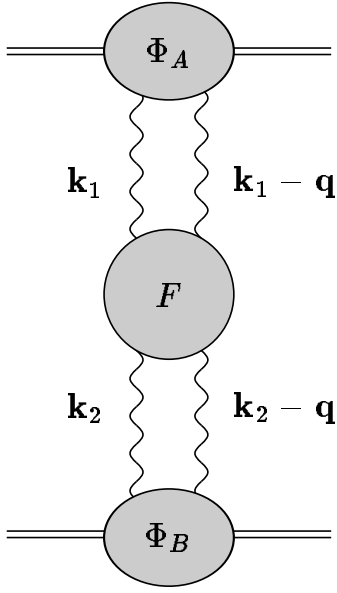}}
\end{center}
\caption{BFKL Pomeron exchange in elastic scattering.}
\label{fig1}
\end{figure}

We start by recalling the results of Lipatov \cite{lip} for the solution of the
BFKL equation for arbitrary momentum transfer, $\q/^2 = -t$. The Pomeron
is described by the universal four-point function, $F$, 
depicted in Fig.1.
In terms of this function, the amplitude for scattering the colourless
states $A$ and $B$ elastically is  
\begin{equation}
 A(s,t) = \frac{i s}{(2 \pi)^4}
\int d^2\ka/ \; d^2\kb/ \; \;
\Phi_A(\ka/,\q/) \;
F(y,\ka/,\kb/,\q/)\;
\Phi_B(\kb/,\q/) \; .
\label{eqng}
\end{equation}
The impact factors $\Phi_A$ and $\Phi_B$ determine the coupling of two
gluons to the external states and we define them to contain no propagator
factors (these are contained in $F$).
Rather than work in transverse momentum space, it is more
convenient to work in the space of impact parameters. Accordingly we define
\begin{eqnarray}
F(y,\ka/,\kb/,\q/)
&=& {1 \over (2 \pi)^6} 
\int d^2 \ba/ \; d^2\bap/ \; d^2\bb/ \; d^2\bbp/ \; \left \{
e^{-i[\ka/.\ba/+(\q/-\ka/).\bap/-\kb/.\bb/-(\q/-\kb/).\bbp/]} 
\right . \nonumber \\
& & \left . f(y,\ba/,\bap/,\bb/,\bbp/) \right \}.
\label{eqnb}
\end{eqnarray}
For future notational convenience we denote this transformation 
\begin{equation}
 F(y,\ka/,\kb/,\q/) = 
\T/ \left \{ f(y,\ba/,\bap/,\bb/,\bbp/) \right \}.
\end{equation}
Lipatov determined that
\begin{equation}
f(y,\ba/,\bap/,\bb/,\bbp/) = 
\int_{-\infty}^{+\infty}
d \nu \; {\nu^2 \over (\nu^2+1/4)^2} \; E^{\nu}(\ba/,\bap/) \; 
E^{\nu*}(\bb/,\bbp/) \; e^{\alp/ \chi(\nu) y}\; 
\label{eqnc}
\end{equation}
where $\alp/ = N_c \alpha_s/\pi$ and the function $\chi(\nu)$ is given by
\begin{equation}
\chi(\nu)= 2 \psi(1)-\psi(1/2+i \nu)-\psi(1/2-i \nu) \; .
\label{eqnd}
\end{equation}
Here $\psi(z)$ is the 
logarithmic derivative of the gamma function, $\Gamma(z)$.
The eigenfunctions, $E^\nu$, are given by
\begin{equation}
E^{\nu}(\ba/,\bap/) = 
\left ( {|\ba/-\bap/| \over |\ba/| |\bap/|} \right )^{1+2 i \nu}.
\label{eqnf}
\end{equation}
We have not included the contributions from non-zero 
conformal spin, since these vanish in the high energy limit. 

\begin{figure}
\begin{center}
\leavevmode
\hbox{\epsfxsize=2.5 in 
\epsfbox{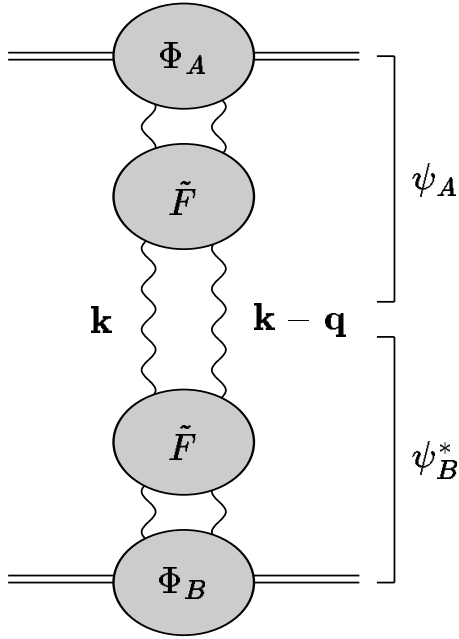}}
\end{center}
\caption{Elastic scattering amplitude as a convolution of the
wavefunctions $\psi_A$ and $\psi_B^*$.}
\label{fig2}
\end{figure}

It is our aim to study the internal dynamics of the BFKL Pomeron.
To do this we ``break'' the Pomeron into two pieces, as shown in Fig.2.
The procedure requires a
little care, in order to account for the propagators which are
present on the external legs. We utilize the following identity:
\begin{equation}
\F/(y,k_1,k_2,q) = \int d^2\k/ \; \F/(y',k_1,k,q) \; 
\F/(y-y',k,k_2,q)
\end{equation}
where
\begin{equation}
\F/(y,k_1,k_2,q) 
= {k_1 (k_1-q) k_2^* (k_2-q)^*} \; F(y,\ka/,\kb/,\q/).
\end{equation}
We have introduced complex numbers to represent the transverse momentum
vectors ($k = k_x + i k_y$). This definition means that $\F/$ 
contains the part propagators $1/k_1^*$, $1/(k_1-q)^*$, $1/k_2$ and 
$1/(k_2-q)$ on its external legs.
The wavefunctions, $\psi_{A,B}$, contain the BFKL dynamics and are given by
\begin{equation}
\psi_A(y,\k/,\q/) = \int d^2\ka/ \; 
\Phi_A(\ka/,\q/) \;  
{\F/(y,k_1,k,q) \over k_1 (k_1-q)} \; \label{psia}
\end{equation}
and
\begin{equation}
\psi^*_B(y,\k/,\q/) = \int d^2\kb/ \; 
\Phi_B(\kb/,\q/) \; 
{\F/(y,k,k_2,q) \over k_2^* (k_2-q)^*}.
\end{equation}
The scattering amplitude is then given by
\begin{equation}
A(s,t)= \frac{i s}{(2 \pi)^4}
\int d^2 \k/ \; \; \psi_A(y',\k/,\q/) \; \psi^*_B(y-y',\k/,\q/). 
\end{equation} 
In terms of Lipatov's solution in impact parameter space we have
\begin{equation} 
\frac{\F/(y,k_1,k_2,q)}{k_1 (k_1-q)} = 
\T/ \left\{
(4 \partial_{b_2} \partial_{b_2'}) 
f(y,b_1,b_1',b_2,b_2') \right\}.
\end{equation}
Similarly, 
\begin{equation}
{\F/(y,k_1,k_2,q) \over k_2^* (k_2-q)^*} = 
\T/ \left \{
(4 \partial_{b_1^*} \partial_{b_1'^*}) 
f(y,b_1,b_1',b_2,b_2') \right \}.
\end{equation}
We can easily calculate 
$\partial_{b_2} \partial_{b_2'}
f(y,b_1,b_1',b_2,b_2')$
using  
\begin{equation}
\partial_{b_2} \partial_{b_2'}
E^{\nu*}(\bb/,\bbp/) =
{(\nu^2+1/4) \over (b_2-b_2')^2} \; E^{\nu*}(\bb/,\bbp/).
\end{equation}
We thus have
\begin{equation}
{\F/(y,k_1,k_2,q) \over k_1 (k_1-q)} = 
\T/ \left \{
\int_{-\infty}^{+\infty}
d \nu \; {4 \nu^2 \over (\nu^2+1/4)} \; \frac{ E^{\nu}(\ba/,\bap/) \; 
E^{\nu*}(\bb/,\bbp/)}{(b_2-b_2')^2} \;  e^{\alp/ \chi(\nu) y} 
\right \} \; . \label{ftil}
\label{eqnhh}
\end{equation}

For our purposes it is beneficial to work in terms of the 
momentum transfer, $\q/$, and size, $\b/$.
Transforming to this mixed representation, we define the new wavefunctions:
\begin{equation}    
\Psi_A(y,\b/,\q/) = \int {d^2\k/ \over 2 \pi } \;
e^{-i\k/.\b/} \; \psi_A(y,\k/,\q/) \;
\label{eqni}
\end{equation}
and
\begin{equation}
\Psi^*_B(y,\b/,\q/) = \int {d^2\k/ \over 2 \pi } \;
e^{+i\k/.\b/} \; \psi^*_B(y,\k/,\q/) \; .
\label{eqnib}
\end{equation}
The scattering amplitude is now given by
\begin{equation}
A(s,t) = \frac{is}{(2 \pi)^4}
\int d^2 \b/ \; \Psi_A(y',\b/,\q/) \; \Psi^*_B(y-y',\b/,\q/). 
\label{eqnj}
\end{equation}
By evaluating these wavefunctions we are able to compute the
probability for finding the Pomeron with size $\b/$ at rapidity $y'$.

To this end, we shall proceed to compute the wavefunction, $\Psi_A$. 
Using (\ref{psia},\ref{ftil},\ref{eqni}) we obtain
\begin{equation}
\Psi_A(y,\b/,\q/) = {1 \over (2 \pi)^6} 
\int d \nu \; {4 \nu^2 \over (\nu^2 + 1/4)} 
\; e^{\alp/ \chi(\nu) y} \; 
V^A_\nu(\q/) \; W_\nu(\b/,\q/)
\label{eqnr}
\end{equation}
where 
\begin{equation}
V^A_\nu(\q/)= \int d^2 \ka/ \; d^2 \ba/ \; d^2 \bap/ \;
e^{-i(\ka/.\ba/+(\q/-\ka/).\bap/)} \;
\Phi_A(\ka/,\q/) \;
E^{\nu}(\ba/,\bap/) \label{eqns}
\end{equation}
and
\begin{equation}
W_\nu(\b/,\q/)= \int {d^2 \k/ \over 2 \pi} \; 
d^2 \bb/ \; d^2 \bbp/ \;
e^{-i(\k/.\b/-\k/.\bb/-(\q/-\k/).\bbp/)} \;
{E^{\nu*}(\bb/,\bbp/) \over (b_2-b_2')^2} \; . \label{wint}
\end{equation}

All of the dependence on the impact parameter $\Phi_A(\ka/,\q/)$
is contained within the function $V^A_\nu(\q/)$.
We proceed to calculate
$W_\nu(\b/,\q/)$ and $V^A_\nu(\q/)$ separately.
The $\k/$ integral of (\ref{wint}) produces a delta function which fixes
$\b/=\bb/-\bbp/$.
Introducing $\R/=(\bb/+\bbp/)/2$ we obtain
\begin{equation}
W_{\nu}(\b/,\q/)=  \frac{2\pi}{b^2} \int d^2 \R/ \;
e^{+i\q/.(\R/-\b//2)} \;
\left ( 
{ \b/^2 \over (\R/-\b//2)^2 \; (\R/+\b//2)^2}
\right )^{1/2-i\nu}. 
\end{equation}
The integral over $\R/$ can be performed using standard Feynman
parametrization techniques: 
\begin{equation}
W_{\nu}(\b/,\q/)= 
{(2 \pi)^2 \mb/ \over \Gamma^2(1/2-i \nu) \; b^2 }
\left ( {\q/^2 \over 4} \right )^{-i \nu}
\int_0^1 dx 
{ e^{-i\q/.\b/x} \over \sqrt{x(1-x)}} \;
K_{2 i \nu}(|\q/|\mb/\sqrt{x(1-x)})
\label{eqnv}
\end{equation}
where $K_{2i\nu}$ is a modified Bessel function. 
The $x$ integral can be performed exactly. In particular, we note that
\begin{equation}
\int_0^1 dx \frac{e^{-i\q/.\b/x}}{\sqrt{x(1-x)}} 
K_{2 i \nu}(|\q/|\mb/\sqrt{x(1-x)}) = e^{-i\q/.\b/ /2} \int_0^1 \frac{2 dx}{
\sqrt{1-x^2}} \cos(\frac{1}{2} \q/.\b/ \sqrt{1-x^2}) 
K_{2i \nu}(\frac{1}{2} |\q/|\mb/ x) \label{int}
\end{equation}
and that the right hand side is tabulated in Gradshteyn \& Ryzhik \cite{gr}
(6.737(3)). We find
\begin{equation}
W_{\nu}(\b/,\q/) = {2i \pi^4 \mb/ \over \Gamma^2(1/2-i \nu) \; b^2}
\left ( {\q/^2 \over 4} \right )^{-i \nu} e^{-i\frac{\q/.\b/}{2}} 
\Delta_{\nu}(\b/,\q/)
\label{eqnbbb}
\end{equation}
where
\begin{eqnarray}
\Delta_{\nu}(\b/,\q/) &=& \frac{1}{{\rm sinh} 2 \pi \nu} \left[ 
e^{\pi \nu} J_{i \nu}\left( \frac{\sqrt{(\q/.\b/)^2 -
\q/^2 \mb/^2} - \q/.\b/}{4} \right)
J_{i \nu}\left( \frac{\sqrt{(\q/.\b/)^2 -
\q/^2 \mb/^2} + \q/.\b/}{4} \right) \right. \nonumber \\ &-&  
(\nu \to -\nu)\Bigg].
\end{eqnarray}
Equivalently\footnote{We note that this is
consistent with the formula given in \cite{np}.}, we have
\begin{equation}
\Delta_{\nu}(\b/,\q/) = \frac{2 i}{{\rm sinh} 2 \pi \nu}
{\rm Im} \left[ J_{i \nu}(b q^*/4) J_{i \nu}(b^* q/4) \right].
\end{equation}

We next turn our attention to $V^A_\nu(\q/)$.
This requires some choice for the impact factor $\Phi_A(\ka/,\q/)$. 
We choose to write the impact factor in the form
\begin{equation}
\Phi_A(\k/,\q/)=
\int \; d^2\r/ \; f_A(\r/) \; 
\left ( e^{i\k/.\r//2}-e^{-i\k/.\r//2}  \right )   
\left ( e^{i(\q/-\k/).\r//2}-e^{-i(\q/-\k/).\r//2} \right ).   
\end{equation}
This is a general choice, appropriate for scattering off colourless
states. 
Assuming that $f_A(\r/) = f_A(|\r/|)$, the calculation of the function 
$V^A_\nu(\q/)$ can be computed using the
method demonstrated in \cite{bflw}. The result is
\begin{eqnarray} 
V^A_\nu(\q/)&=&
2 (2 \pi)^5 \; 
{\Gamma(1/2- i \nu) \over \Gamma(1/2 + i \nu)}
\left ( \q/^2 \over 4 \right ) ^{-1+i \nu}
\int_{1/2-i \infty}^{1/2+i \infty} {dz \over {2 \pi i}} \;
\left \{ 
\int dr \; \left ( \frac{\q/^2 \r/^2}{16} \right )^z \; 
(-f_A(\r/)) \right . \; \nonumber \\
&& \;
{\Gamma(1-z-i \nu) \over {\Gamma(1/2+z/2-i \nu/2) \; 
\Gamma(1-z/2-i \nu/2)}} \; \nonumber \\
&& \; \left. 
{\Gamma(1-z+i \nu) \over {\Gamma(1/2+z/2+i \nu/2) \;
\Gamma(1-z/2+i \nu/2)}} \right\}.
\label{eqnzc}
\end{eqnarray}
This is a particularly convenient way of expressing $V^A_\nu$ since it
allows the limits $1/\q/^2 \gg \r/^2$ and $1/\q/^2 \ll \r/^2$ to be
extracted with ease (by focusing on the poles lying closest to the
contour). Alternatively, we could utilize (\ref{int}) to write
\begin{equation}
V^A_\nu(\q/) =  8i \pi^5 \; \int d^2 \r/ \; f_A(\r/) \; |\r/| e^{-i \q/.\r/}
\left( \frac{ \q/^2}{4} \right)^{i \nu} \Delta_\nu(\r/,\q/).
\end{equation}
This latter form does not assume $f_A(\r/) = f_A(|\r/|)$.


\vspace*{4mm}
\noindent {\large \bf 3. Inside the BFKL Pomeron}

We are now in a position to investigate the typical distances involved
in the exchange. We can ``look inside'' the Pomeron and examine 
its size at some intermediate rapidity, $y'$. The probability density
for having a Pomeron with size $\mb/$ at rapidity $y'$ is proportional
to the product of the wavefunctions, $\Psi_A(y',\b/,\q/) 
\Psi_B^*(y-y',\b/,\q/)$ (see (\ref{eqnj})).
We need to make some assumption regarding the nature of the external
states in order to deduce the process dependent $V^A_\nu$.
We work with (\ref{eqnzc}) under the assumption 
that the external state is
characterised by a single scale, $Q$. In this case, we can write
\begin{equation}
\int_0^\infty dr \left( \frac{\q/^2 r^2}{16} \right)^z (-f(r)) =
Q \left( \frac{\q/^2}{16 Q^2} \right)^z h(z)
\end{equation}
where $h(z)$ is some dimensionless function of $z$ which contains poles
only to the left of the $z$ plane contour, Re $z =
1/2$, otherwise the impact factor $\Phi_A$ is not defined. 
For photon and vector meson external states the following
expressions for $h(z)$ are needed:
\begin{eqnarray}
-f(r) = Q^2 K_0(Q r) \hspace*{3cm} & \Rightarrow 
& h(z) = 2^{2z-1} \Gamma^2(1/2+z) \nonumber \\
-f(r) = Q^2 K_1(Q r) \hspace*{3cm} & \Rightarrow
& h(z) = 2^{2z-1} \Gamma(1+z) \Gamma(z).
\end{eqnarray}
By way of example we also note that
\begin{eqnarray}
-f(r) = \frac{Q}{r} e^{-Q^2 r^2} \hspace*{3cm} & \Rightarrow
& h(z) = \frac{\Gamma(z)}{2} \nonumber \\
-f(r) = r Q^3 e^{-Q^2 r^2} \hspace*{3cm} & \Rightarrow
& h(z) = \frac{\Gamma(1+z)}{2} \nonumber \\
-f(r) = \delta(r^2 - 1/Q^2) \hspace*{3cm} & \Rightarrow
& h(z) = 1/2.
\end{eqnarray}
The existence of double poles induces additional logarithms in the final
answer but does not affect any of our main conclusions. 
Subsequently, we will therefore assume that $h(z)$ contains only single poles.
We can now proceed to examine the wavefunction in the limits 
$\q/^2 \ll Q^2 $ and $\q/^2 \gg Q^2$. 

\vspace*{4mm}
\noindent {\large \bf The case when $\q/^2 < 16 \; Q^2$} \\
In this case 
the $z$-plane contour of (\ref{eqnzc}) can be closed in the right half plane.
We consequently pick up the poles in $V_\nu^A$ which come from the process
independent part, i.e. at $z = 1 \pm i \nu$. Evaluating the integral
keeping only the nearest pole gives
\begin{eqnarray}
V_\nu (\q/) &=& \frac{2(2 \pi)^5 Q}{\sqrt \pi}  
{\Gamma(1/2-i\nu) \over \Gamma(1/2+i\nu)} 
\left( \frac{\q/^2}{4} \right ) ^{-1+i \nu}
\left [ 
\left ( \frac{\q/^2}{16 Q^2} \right )^{1+i \nu} 
{h(1+i \nu) \Gamma(-2 i\nu) \over \Gamma(1/2-i\nu) 
\Gamma(1+i\nu)} \right . \nonumber \\
&& \; \; \; \; \left .
+ \left ( \frac{\q/^2}{16 Q^2} \right )^{1-i \nu} 
{h(1-i \nu) \Gamma(2i\nu) \over \Gamma(1/2+i\nu) 
\Gamma(1-i\nu)}
\right ] \; .
\label{eqnzdd}
\end{eqnarray}
The corrections to this expression are suppressed by powers of $\q/^2/Q^2$.
We have only the $\nu$ integral to perform in order to obtain the
wavefunction $\Psi_A$. In order to proceed further 
we consider the two regions $\vert \q/ \vert > 1/ \mb/$ and $|\q/| < 1/\mb/$ 
separately.

In the region where $|\q/|\mb/ \lapproxeq 1$
we can use the small argument expansions of the Bessel functions, and
for large $y$ the dominant contribution to the $\nu$ integral 
in (\ref{eqnr}) will come from the region of small $\nu$, hence we may write
\begin{equation}
{W}_\nu(\b/,\q/) \approx
{(2 \pi)^2 \mb/ \over 2 \nu b^2}
\left ( {\q/^2 \over 4} \right )^{-i \nu}
\sin \left ( \nu \ln(16 /(\q/^2 \b/^2))  \right ).
\label{eqny}
\end{equation}
Whilst for $V_\nu$ the corresponding small $\nu$ limit of (\ref{eqnzdd}) is
\begin{equation}
V_\nu^A(\q/) \approx { 16 h(1) \pi^4 \over Q \nu}
\left ( \q/^2 \over 4 \right ) ^{i \nu}
\sin \left ( \nu \ln(16 Q^2/\q/^2) \right ) \; .
\label{eqnzee}
\end{equation}
For the wavefunction, we thus obtain
\begin{equation}
\Psi_A(y,\b/,\q/) \approx 
8 h(1) \frac{\mb/}{ Q b^2}
\int d \nu \;   e^{\alp/ \chi(\nu) y} \; 
\sin \left ( \nu \ln(16/(\q/^2 b^2)) \right) \;
\sin \left ( \nu \ln(16 Q^2/\q/^2 \right) .
\end{equation}
Expanding the BFKL eigenfunction about $\nu=0$ allows the integral to be
performed.
Using $\chi(\nu) \approx 4 \ln 2 - 14 \zeta(3) \nu^2$
yields
\begin{equation}
{\Psi}_A(y,\b/,\q/)=
4 \sqrt{\pi} h(1) \frac{\mb/}{Q b^2}
{e^{\omega_0 y} \over (a^2 y)^{1/2}} \;
\left [
\exp \left ( {-\ln^2(\b/^2 Q^2) \over 4 a^2 y} \right )
-
\exp \left ( {-\ln^2(256 Q^2/(\q/^4 \b/^2)) \over 4 a^2 y} \right )
\right ].
\label{eqnzgg}
\end{equation}
Here $a^2 = 14 \alp/ \zeta(3)$.

In the region where $|\q/|\mb/ \gg 1$ the function 
$V_\nu^A$ is again given by
(\ref{eqnzee}).
For $W_\nu$ we take the large argument approximation to the
Bessel functions. After some algebra we find that
\begin{equation}
{W}_\nu(\b/,\q/) \simeq
{(2 \pi)^2  \over |\q/| b^2}
\left ( {\q/^2 \over 4} \right )^{-i \nu}
\; (1+e^{-i\q/.\b/}) \; .
\label{eqnza}
\end{equation}
Inserting (\ref{eqnzee}) and (\ref{eqnza}) 
into our expression for $\Psi_A$ (\ref{eqnr}) yields
\begin{equation}
{\Psi}_A(y,\b/,\q/) \approx
{16 h(1) \over Q |\q/| b^2} (1+e^{-i\q/.\b/})
\int d\nu \;
e^{\alp/ \chi(\nu) y} \;
\nu \sin \left (\nu \ln(16 Q^2/\q/^2) \right) 
\end{equation}
which evaluates (on expanding the eigenvalue) to
\begin{equation}
{\Psi}_A(y,\b/,\q/)=
{8 h(1) \sqrt{\pi} \over Q |\q/| b^2} (1+e^{-i\q/.\b/})
{e^{\omega_0 y} \over (a^2 y)^{3/2}}
\ln(16 Q^2 /\q/^2) 
\exp(-\ln^2(16 Q^2/\q/^2)/4a^2 y).  \label{psif2}
\end{equation}

Note that in the limit $\q/ \to 0$ we only have the region described
by (\ref{eqnzgg}) in which the second exponential term vanishes. This
corresponds to the well known result for diffusion about
the size of the external state, $Q$. 
In that case there is nothing to 
prevent diffusion into the large distance region where a
perturbative calculation may be unreliable.
(see, e.g. \cite{bart} for a
more detailed study of diffusion at $t=0$.)
However, the effect of a finite $\q/^2$ is dramatic. 
Equation (\ref{psif2}) reveals that there is no diffusion 
inside the region $|\b/| |\q/| > 1$. 
Moreover, the distribution in this region 
vanishes rapidly as $\b/$ increases.
The result is that diffusion to 
large sizes is effectively blocked beyond the scale 
$1/ \vert \q/ \vert$ \cite{lotter}. 
The momentum transfer acts as an effective infrared cut-off. 
In Fig.~3  we illustrate 
the diffusion for the case $\q/=0$.
This should be compared with Fig.~4 in which 
we show the equivalent distributions for the case $\q/^2= 1$ GeV$^2$.
In generating these plots we chose a simple Gaussian form for the
function $f(r)$ which characterises the external impact factor,
$|f(r)| = Q/r e^{-Q^2 r^2}$. 
As we noted earlier this corresponds to a function $h(z) = \Gamma(z)/2$. 
\begin{figure}[h]
\begin{center}
\leavevmode
\hbox{\epsfxsize=6.5 in 
\epsfbox{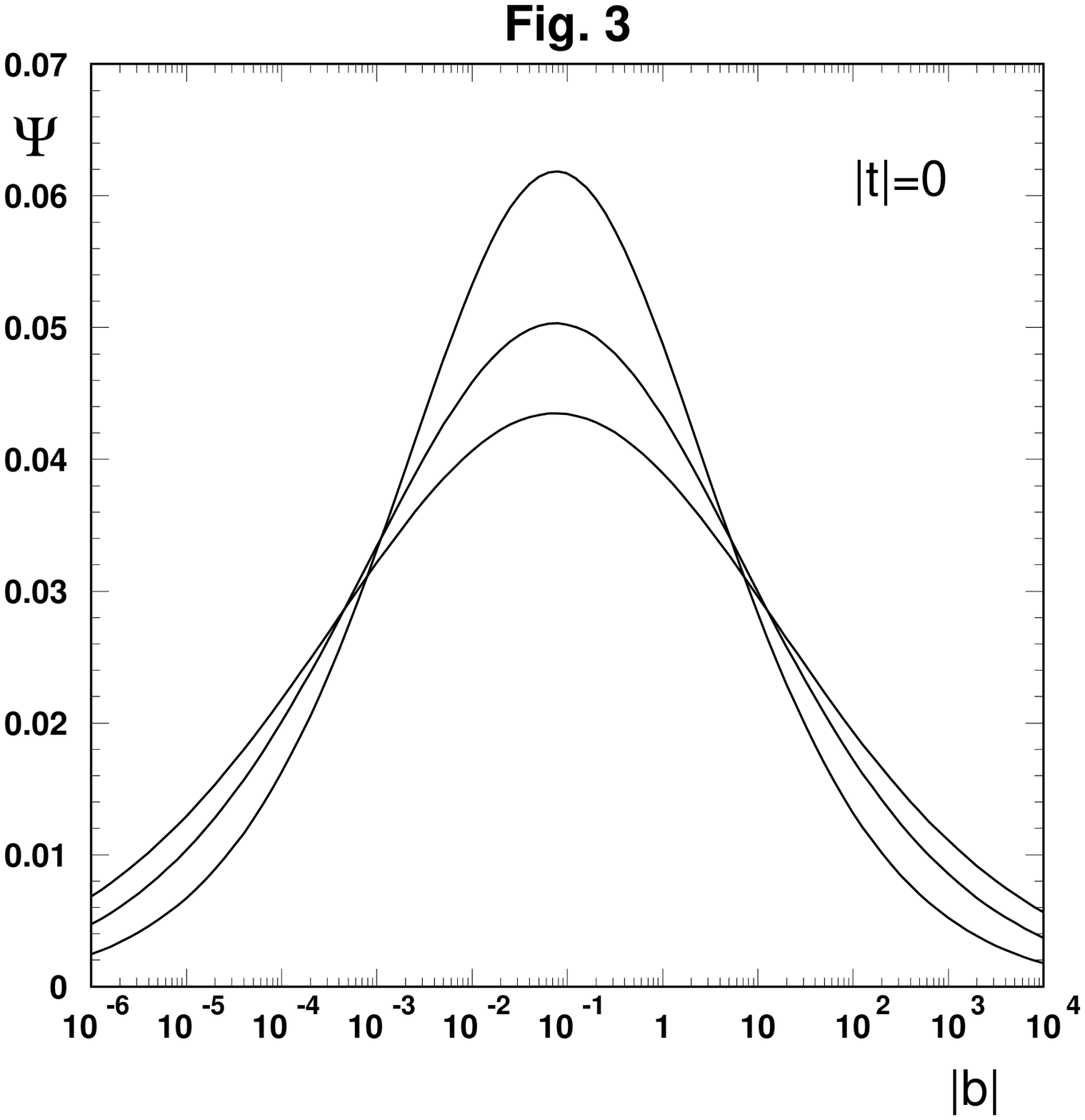}}
\end{center}
\caption[junk]{The distribution of gluon size inside the
BFKL Pomeron for $|t|=\q/^2=0$ and $Q=10$ GeV 
as described by $ \Psi = \Psi_A \; b^2/\mb/$. The plot
shows the broadening of the distribution as the rapidity, $y$, is
increased through the values $y=10,15$ and 20.
Nothing prevents diffusion
into the long distance region.
The curves are from a
numerical calculation of (\ref{eqnr}) using 
(\ref{eqnbbb}) and (\ref{eqnzc}). 
Each curve has been scaled by a factor $e^{-\omega_0 y}$.}
\label{fig3}
\end{figure}
\begin{figure}[h]
\begin{center}
\leavevmode
\hbox{\epsfxsize=6.5 in 
\epsfbox{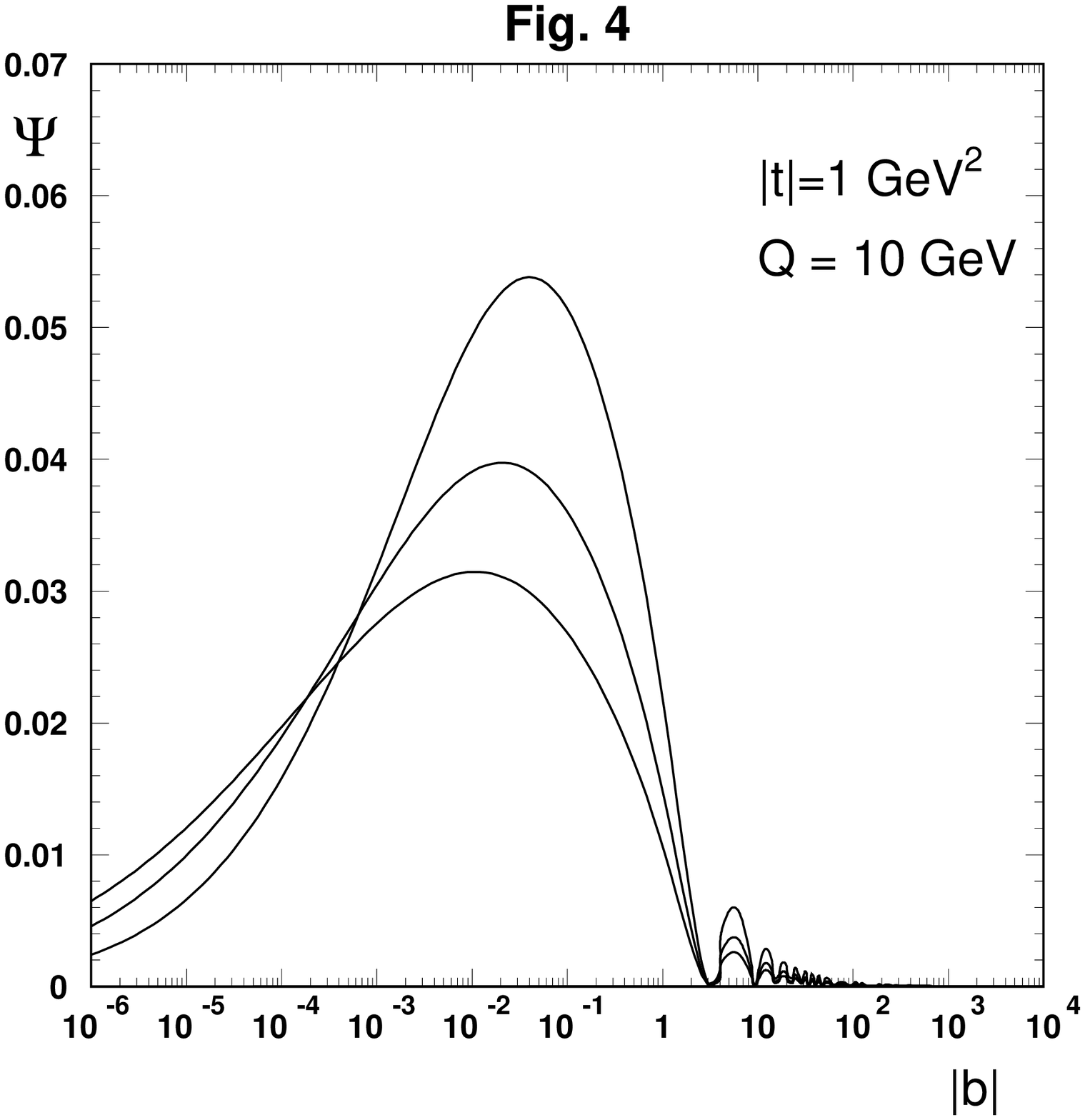}}
\end{center}
\caption[junk]{The distribution of gluon size 
inside the
BFKL Pomeron for $|t|=\q/^2=1$ GeV$^2$ and $Q=10$ GeV 
as described by $ \Psi = \Psi_A \; b^2/\mb/$. The plot
shows the broadening of the distribution as the rapidity, $y$, 
is increased through the values $y=10,15$ and 20. 
The scale $\q/$ acts as an effective infrared cut off preventing
diffusion into the infrared region. 
The curves are from a
numerical calculation of (\ref{eqnr}) using 
(\ref{eqnbbb}) and (\ref{eqnzc}). 
Each curve has been scaled by a factor $e^{-\omega_0 y}$.}
\label{fig4}
\end{figure}

\vspace*{4mm}
\noindent {\large \bf The case when $\q/^2 > 16 \; Q^2$}

For this case the $z$-plane contour integral must be closed in the left half
plane. The only poles are those which arise from the process dependent
factor $h(z)$. In the limit $\q/^2 \gg 16 \; Q^2$ the rightmost pole gives
the leading power dependence. If this is a single pole at $z=-z_0$ then
we can take
\begin{equation}
h(z) \approx \frac{h_L(z)}{z+z_0}.
\end{equation}
We can now evaluate $V_\nu^A$:
\begin{eqnarray}
V_\nu (\q/) &=& 2 (2 \pi)^5  
{\Gamma(1/2-i\nu) \over \Gamma(1/2+i\nu)} 
\left( \frac{\q/^2}{4} \right ) ^{-1+i \nu} Q h_L(-z_0)
\left ( \frac{16 Q^2}{\q/^2} \right )^{z_0} \nonumber \\
&& \;
{\Gamma(1-z_0-i \nu) \over {\Gamma(1/2+z_0/2-i \nu/2) \; 
\Gamma(1-z_0/2-i \nu/2)}} \; \nonumber \\
&& \;
{\Gamma(1-z_0+i \nu) \over {\Gamma(1/2+z_0/2+i \nu/2) \;
\Gamma(1-z_0/2+i \nu/2)}}. 
\label{eqnzx}
\end{eqnarray}
The corrections are suppressed by powers of $Q^2/\q/^2$. 
We can again separate the analysis into two regions consisting of
large and small $|\q/| |\b/|$.

In the region where $|\q/|\mb/ \lapproxeq 1$
the function ${W}_\nu(\b/,\q/)$ is again given by (\ref{eqny}).
For $V_\nu$ the corresponding small $\nu$ limit of (\ref{eqnzx}) is
\begin{equation}
V_\nu^A(\q/) \approx 64 \pi^4
\left ( \q/^2 \over 4 \right )^{-1+i \nu}
\left( \frac{64 Q^2}{\q/^2}\right)^{z_0} Q \; \tilde{h}_L(z_0) \; . 
\label{eqnze}
\end{equation}
where we have further defined
\begin{equation}
\tilde{h}_L(z_0) = h_L(-z_0) \left(
\frac{\Gamma(1/2+z_0/2)}{\Gamma(1/2-z_0/2)} 
\right)^2.
\end{equation}
This form is only appropriate providing $z_0$ does not induce poles in
the denominator, i.e. $z_0$ is not an odd integer. 
Although it would be straightforward
to accommodate such values of $z_0$ (since they merely produce an
additional factor of $\nu^2$ in the numerator)
we shall ignore this possibility in the following.
This does not affect
our main conclusions.
For the wavefunction, we now obtain
\begin{equation}
\Psi_A(y,\b/,\q/) \approx 
128 \tilde{h}_L(z_0) \frac{Q \mb/}{b^2 \q/^2} \left( \frac{64
Q^2}{\q/^2} \right)^{z_0}
\int d \nu \; \nu \;  e^{\alp/ \chi(\nu) y} \; 
\sin\left( \nu \ln(16/(\q/^2 \b/^2)) \right).
\end{equation}
Expanding the BFKL eigenfunction about $\nu=0$ gives
\begin{equation}
{\Psi}_A(y,\b/,\q/)=
64 \sqrt{\pi}\tilde{h}_L(z_0) \frac{Q \mb/}{b^2 \q/^2}
\left( \frac{64 Q^2}{\q/^2} \right)^{z_0}
{e^{\omega_0 y} \over (a^2 y)^{3/2}} \ln(16/(\q/^2 \b/^2))
\exp \left ( {-\ln^2(16/(\q/^2 \b/^2)) \over 4 a^2 y} \right ).
\label{eqnzg}
\end{equation}
In the region where $|\q/|\mb/ \gg 1$ the function 
$V_\nu^A$ is again given by
(\ref{eqnze}) whilst 
${W}_\nu(\b/,\q/)$ is given by (\ref{eqnza}).
Inserting (\ref{eqnze}) and (\ref{eqnza}) 
into our expression for $\Psi_A$ (\ref{eqnr}) yields
\begin{equation}
{\Psi}_A(y,\b/,\q/) \approx
256 {Q \over |\q/|^3 b^2} \tilde{h}_L(z_0)
\left( \frac{64 Q^2}{\q/^2} \right)^{z_0} 
(1+e^{-i\q/.\b/})
\int d\nu \; \nu^2 \; e^{\alp/ \chi(\nu) y}  
\end{equation}
which evaluates (on expanding the eigenvalue) to
\begin{equation}
{\Psi}_A(y,\b/,\q/)=
128 \sqrt{\pi} {Q \over |\q/|^3 b^2} \tilde{h}_L(z_0) 
\left( \frac{64 Q^2}{\q/^2} \right)^{z_0}
(1+e^{-i\q/.\b/})
{e^{\omega_0 y} \over (a^2 y)^{3/2}}.  \label{psif1}
\end{equation}
Notice that the wavefunction of (\ref{eqnzg}) corresponds to diffusion 
about $\b/^2 = 16/\q/^2$, i.e. the diffusion is no longer centred around
the external scale, $Q$, determined by the impact factor.
As before the momentum transfer $\q/$ acts as an effective
infrared cut-off. These properties can be seen in Fig.~5 

\begin{figure}
\begin{center}
\leavevmode
\hbox{\epsfxsize=6.5 in 
\epsfbox{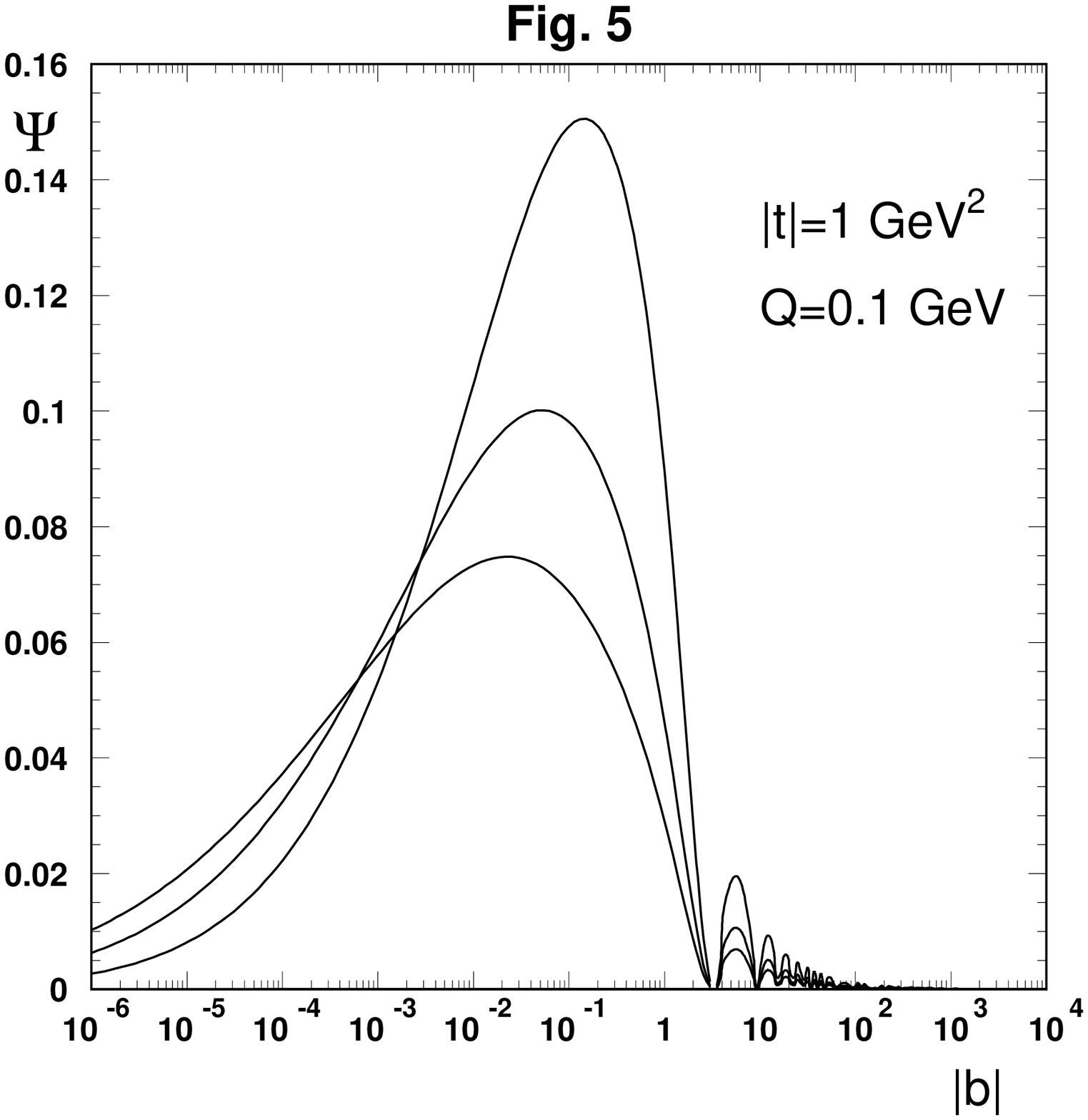}}
\end{center}
\caption[junk]{The distribution of gluon size 
inside the
BFKL Pomeron for $|t|=\q/^2=1$ GeV$^2$ and $Q=0.1$ GeV 
as described by $ \Psi = \Psi_A \; b^2/\mb/$. The plot
shows the broadening of the distribution as the rapidity, $y$, is
increased through the values $y=10,15$ and 20. 
The scale $\q/$ acts as an effective infrared cut off preventing
diffusion into the infrared region. 
The curves are from a full
numerical calculation of (\ref{eqnr}) using 
(\ref{eqnbbb}) and (\ref{eqnzc}). 
Each curve has been scaled by a factor $e^{-\omega_0 y}$.}
\label{fig5}
\end{figure}
\begin{figure}
\begin{center}
\leavevmode
\hbox{\epsfxsize=3.5 in 
\epsfbox{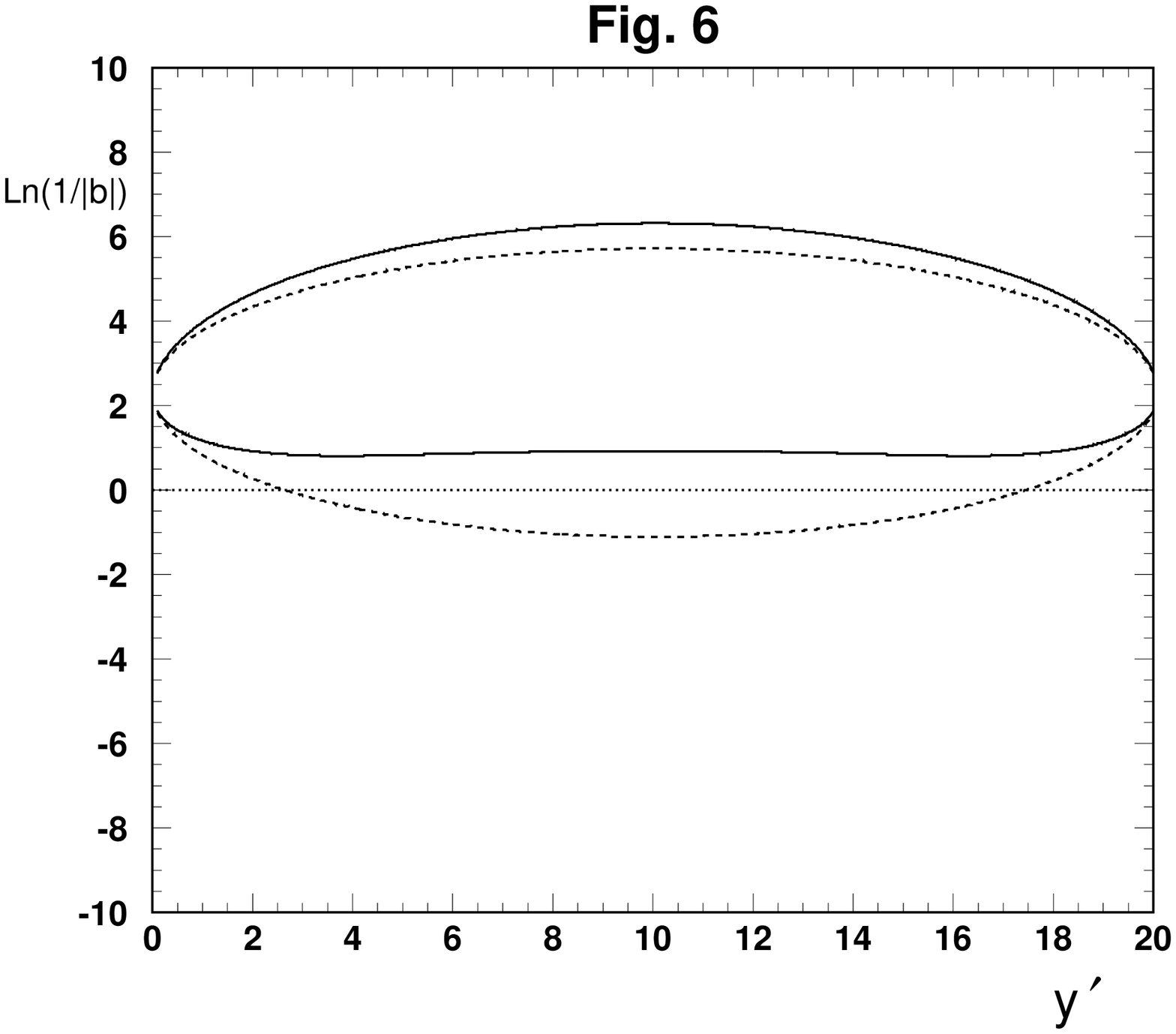}}
\end{center}
\caption[junk]{The width (as defined by
half the maximum) of the distribution 
$\b/^2 \Psi_A \Psi_B^*$ 
for the
case $Q_A = Q_B = 10 $ GeV and $y=20$. 
The dotted lines represent the $|t|=0$ case.
The full lines represent the case 
$|t|=\q/^2=1$ GeV$^2$.}
\label{fig6}
\end{figure}
\begin{figure}
\begin{center}
\leavevmode
\hbox{\epsfxsize=3.5 in 
\epsfbox{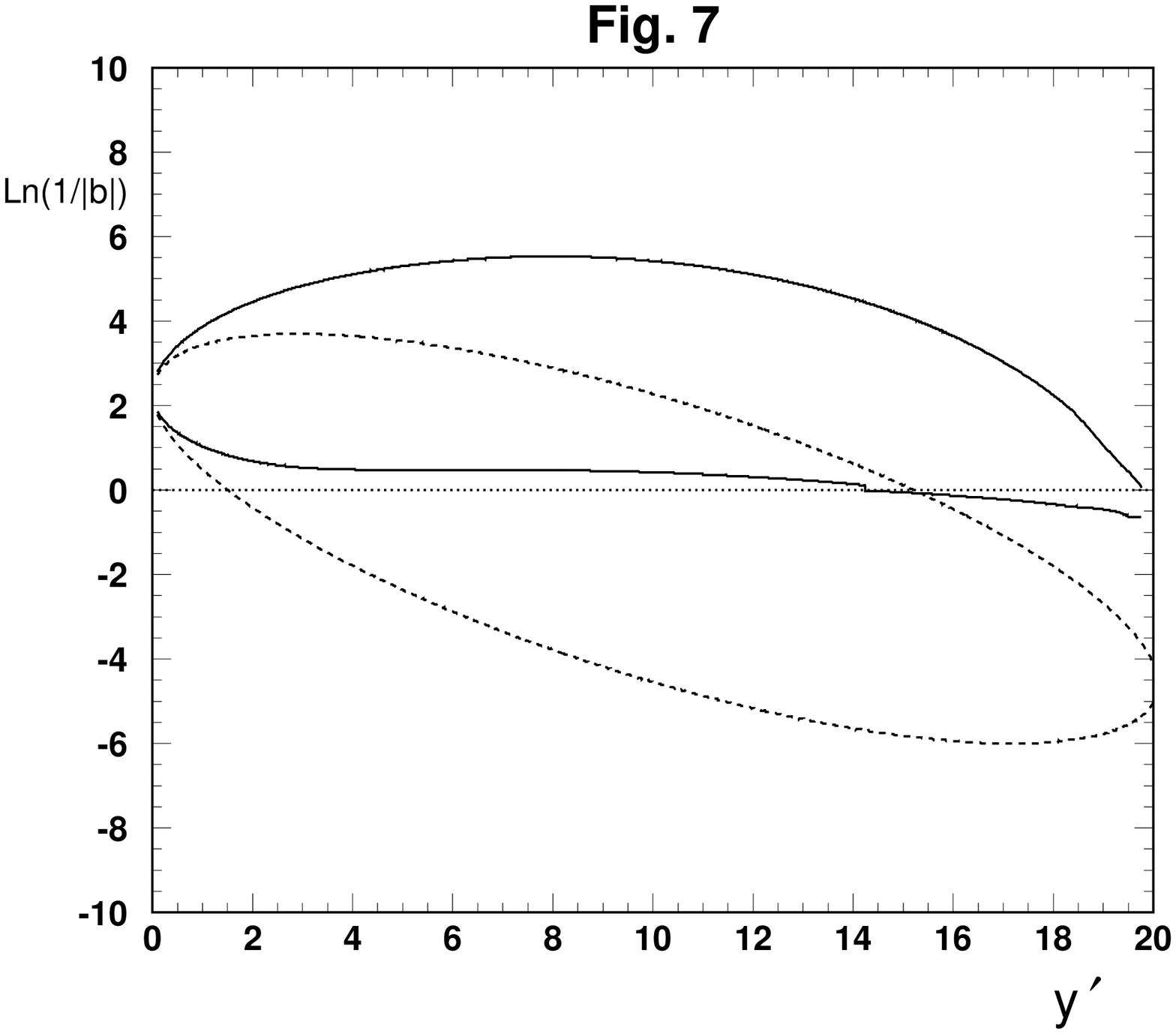}}
\end{center}
\caption[junk]{The width (as defined by
half the maximum) of the distribution 
$\b/^2 \Psi_A \Psi_B^*$ 
for the
case $Q_A = 10 $ GeV, $Q_B = 10^{-2}$ GeV and $y=20$.
The dotted lines represent the $|t|=0$ case.
The full lines represent the case 
$|t|=\q/^2=1$ GeV$^2$.}
\label{fig7}
\end{figure}
\begin{figure}
\begin{center}
\leavevmode
\hbox{\epsfxsize=3.5 in 
\epsfbox{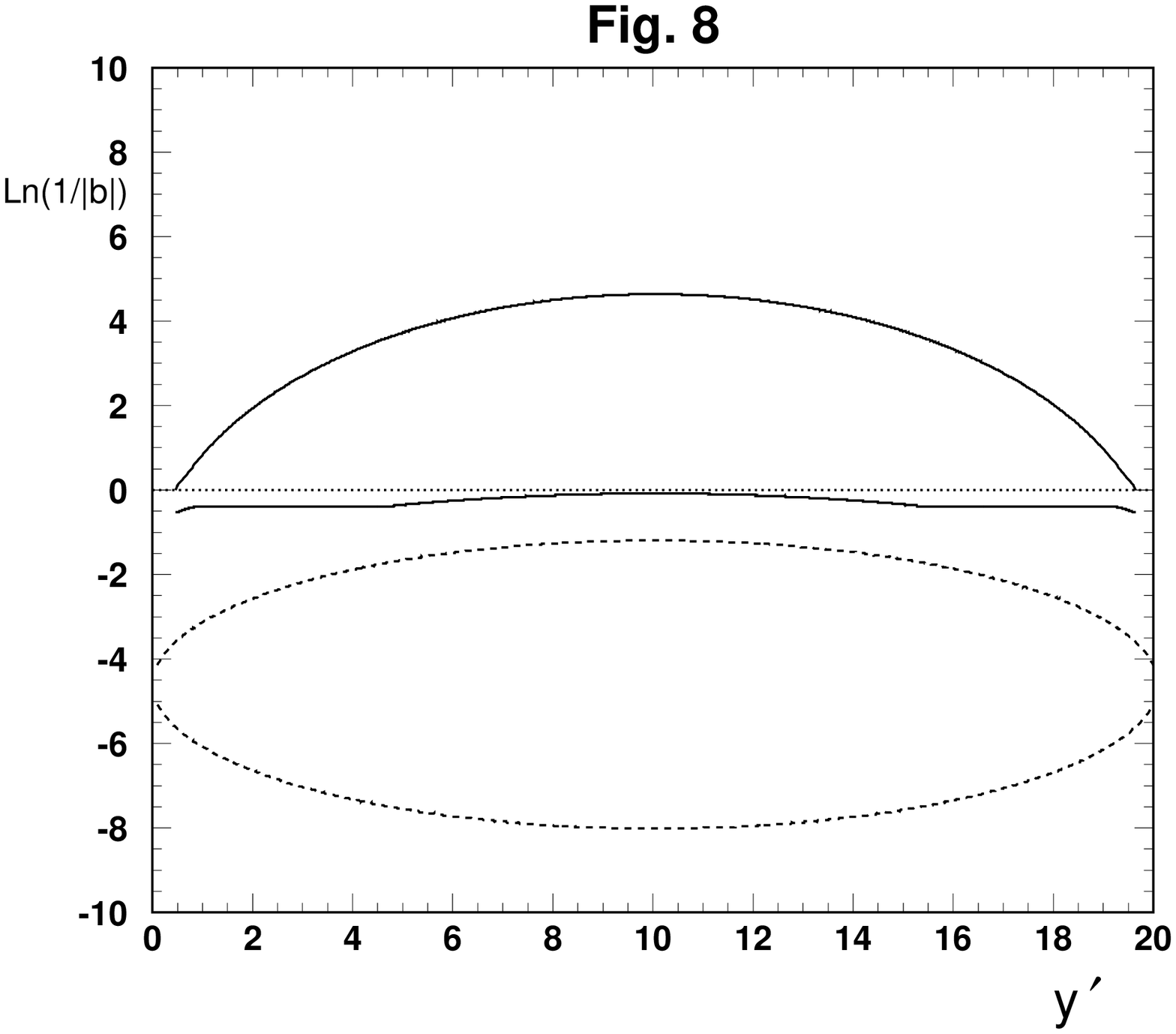}}
\end{center}
\caption[junk]{The width (as defined by
half the maximum) of the distribution 
$\b/^2 \Psi_A \Psi_B^*$ 
for the case $Q_A = Q_B = 10^{-2} $ GeV and $y=20$. 
The dotted lines represent the $|t|=0$ case.
The full lines represent the case 
$|t|=\q/^2=1$ GeV$^2$.}
\label{fig8}
\end{figure}

The wavefunction $\Psi_B^*$ can be computed analogously. For the sake of
illustration, we shall assume that the external states, $A$ and $B$, are
defined in terms of the scales $Q_A$ and $Q_B$ respectively and that in
all other respects they are equal. In this case, $\Psi_B^*$ is simply
the complex conjugate of $\Psi_A$ with $Q_A$ replaced by $Q_B$.
Note that the complex factors in the denominators always combine to
produce the factor $1/\b/^4$ in the probability density ($=\Psi_A
\Psi_B^*$).  We shall evaluate the 
probability distribution $\Psi_A \Psi_B^*$ for three separate cases. 
Firstly, when the external scales ($Q_A$ and $Q_B$) are both larger than
the momentum transfer; secondly, 
when one is large and one is small and finally 
when both are small.
 
In Figures 6,7 and 8, we show the width of the $b$-distribution
as a function of $y'$. The width is
determined by evaluating the value 
of $\ln \b/$ when $\b/^2 \Psi_A \Psi_B^*$
is half of its maximum. Since
\begin{equation}
\frac{A(s,t)}{s} \sim \int d(\ln \b/^2) \;
\b/^2 \; \Psi_A(y',\b/,\q/) \Psi_B^*(y-y',\b/,\q/)
\end{equation}
it follows that for the dominance of short distance physics the
``cigar'' must remain inside the region of small $| \b/ |$. 
Fig.~6 shows the case when both external scales are larger than the 
momentum transfer, $\q/^2$. In this case the diffusion 
is similar to the $\q/^2=0$ case, i.e. increasingly larger distances are
probed as $y'$ approaches $y/2$. However, movement to distances larger than
$\sim 1/|\q/|$ is blocked. This (almost) total exclusion from the region 
$|\b/| \gapproxeq 1/| \q/|$  
remains even if one or both of the external scales
becomes small. In these cases the picture is
quite different from the $\q/^2=0$ case. 
Fig.~7 shows the diffusion when one external scale is large
and the other is small. The case where
both external scales are smaller than
the momentum transfer, $\q/^2$ is shown in Fig.~8.


\vspace*{4mm}
\bigskip
\noindent {\large \bf Acknowledgements:}
We thank Jochen Bartels, Hans Lotter, Douglas Ross, Misha Ryskin and
Mark W\"usthoff for numerous helpful discussions.

\end{document}